\documentclass[aps,prb,twocolumn,superscriptaddress,showpacs]{revtex4-2}

\usepackage[pdftex]{graphicx} 
\usepackage{amsmath} 
\usepackage{amssymb} 
\usepackage{mathtools}
\usepackage{color} 
\usepackage{textcomp} 
\usepackage{lmodern} 
\usepackage[T1]{fontenc}

\begin{document}

\title{Direct measurement of the scattering cross sections of liquid ortho-deuterium for ultracold neutrons and comparison with model calculations} 

\author{Stefan D\"{o}ge}
\email[Corresponding author Stefan Doege: ]{stefan.doege@tum.de}
\affiliation{Institut Laue--Langevin, 71 avenue des Martyrs, F-38042 Grenoble Cedex 9, France}
\affiliation{Physik-Department, Technische Universit\"{a}t M\"{u}nchen, James-Franck-Strasse 1, D-85748 Garching, Germany}

\author{J\"{u}rgen Hingerl}
\affiliation{Institut Laue--Langevin, 71 avenue des Martyrs, F-38042 Grenoble Cedex 9, France}
\affiliation{Physik-Department, Technische Universit\"{a}t M\"{u}nchen, James-Franck-Strasse 1, D-85748 Garching, Germany}

\author{Winfried Petry}
\affiliation{Physik-Department, Technische Universit\"{a}t M\"{u}nchen, James-Franck-Strasse 1, D-85748 Garching, Germany}
\affiliation{Forschungsneutronenquelle Heinz Maier-Leibnitz, Technische Universit\"{a}t M\"{u}nchen, D-85748 Garching, Germany}

\author{Christoph Morkel}
\affiliation{Physik-Department, Technische Universit\"{a}t M\"{u}nchen, James-Franck-Strasse 1, D-85748 Garching, Germany}

\begin{abstract}
Liquid deuterium is a fluid between the quantum and classical regimes. It attracts interest from fundamental research for the verification of quantum calculations but also from neutron physics as it is a widely used neutron moderator medium. We have measured the scattering cross sections of liquid \textit{ortho}-deuterium in the ultracold-neutron range for four different temperatures (19--23~K) and compared them with calculations from a parameter-free analytical calculation model as well as with previous measurements at 19~K by another research group. All three show remarkable agreement, which establishes the validity of the calculation model and proves it is a reliable basis for the derivation of scattering kernels. The deconvolution of our measured transmission data changed the cross section results noticeably only for neutrons faster than 10 m/s. We found the total scattering cross section of liquid deuterium to be inversely proportional to velocity, as is predicted by theory.
\\
\\
Published online on 08 August 2022: \url{https://doi.org/10.1103/PhysRevB.106.054102} \\ S. D\"{o}ge, J. Hingerl, W. Petry, C. Morkel, Physical Review B 106 (5), 054102 (2022) \\ \textcopyright\, 2022. This manuscript version is made available under the \href{https://creativecommons.org/licenses/by/4.0/}{CC-BY 4.0 license}.
\end{abstract}

\maketitle

\section{Introduction}

The hydrogen atom and the molecule diprotium ($^1$H$_2$), with its nuclear-spin interactions and specific heat anomaly at low temperatures~\cite{eucken:1912,dennison:1927}, played a fundamental role in the emergence of quantum mechanics~\cite{heisenberg:1927,hund:1927}. Discovered later, the deuterium atom and the dideuterium molecule ($^2$H$_2$, most often simply called ``deuterium'') served in many respects as systems to check and confirm calculation models for the protium atom and molecule.

Of the two rotational spin species, \textit{ortho}-deuterium is the liquid of interest because the rotational spin equilibrium at temperatures between 19 and 23~K is around 98\% \textit{ortho}-deuterium and only 2\% \textit{para}-deuterium~\cite{farkas:1934,silvera:1980}.

Liquid \textit{ortho}-deuterium (liqD$_2$) has been used extensively as a ``cold source'' at various research centers to moderate thermal neutrons down to cold-neutron energies and to increase the fraction of ultracold neutrons in the spectrum~\cite{ageron:1969,golub:1991}. Recently, it was proposed as a pre-moderator material for novel high-flux converters of ultracold neutrons~\cite{lychagin:2016,schreyer:2020}, which may advance important research projects in particle and astrophysics~\cite{dubbers:2011} and the search for physics beyond the standard model~\cite{dubbers:2011,crivellin:2020}.

In 1970~\cite{seiffert:1970,seiffert:1970-article}, Seiffert published the first experimental scattering data for thermal and cold neutrons in liquid deuterium down to 0.9~meV. Only in 2005, Atchison \textit{et al.} published total cross sections $\sigma_\text{tot}$ for very cold neutrons (VCNs) and ultracold neutrons (UCNs)~\cite{atchison:2005-liq}, and established a questionable velocity dependence of $\sigma_\text{tot}$ according to $\sigma_\text{tot}(v) = \sigma_0 + c/v$, where $\sigma_0$ represents the incoherent elastic scattering cross section of one deuterium molecule. However, liquids do not show elastic scattering~\cite{squires:1978} and for fluids with a Maxwell--Boltzmann velocity distribution, the scattering cross section is proportional to 1/$v$~\cite{lovesey:1984},
\begin{equation}\label{eq:sigma-v}
\sigma_\text{scatt}(v) = c/v,
\end{equation}
where $\sigma_\text{scatt}(v)$ is the scattering cross section of one deuterium molecule in barn at a given velocity, $c$ is a temperature-dependent constant, and $v$ is the neutron's velocity.

Our group also published experimental scattering cross section data for UCNs in 2015~\cite{doege:2015}.

In the same publication, we presented a parameter-free analytical calculation model for the scattering cross sections of liquid deuterium for UCNs and VCNs, which provides a way to calculate the constant $c$ from Eq.~\ref{eq:sigma-v} for given temperatures and \textit{ortho}/\textit{para} ratios.

The aim of this paper is to compare and reconcile all available experimental scattering cross section data with results from model calculations and simulations for the UCN energy range. These consolidated data can then serve to benchmark nuclear data libraries, such as described in a recent review by Plompen \textit{et al.}~\cite{plompen:2020} or the EXFOR database maintained by the International Atomic Energy Agency (IAEA)~\cite{iaea:2014}, \url{https://www-nds.iaea.org/exfor}, which to date include only the ultracold-neutron cross sections measured by Atchison \textit{et al.}~\cite{atchison:2005-liq}.

\section{Theoretical Background}\label{sec:theor-background}

Hamermesh and Schwinger calculated the neutron-scattering cross sections of free deuterons and interaction-free, i.e., gaseous, deuterium molecules in the wake of early neutron research in the 1940s~\cite{hamermesh:1946}. These cross sections were \textit{self} cross sections, neglecting collective interactions in the sample. They were calculated for thermal and subthermal neutrons in deuterium gas at low temperatures. Later, Young and Koppel~\cite{young-koppel:1964} calculated the double differential cross sections of deuterium for the \textit{ortho} and \textit{para} species for a wider temperature and energy range, taking into account rotation, vibration, and translation of the molecule. The special case of a liquid was also described but is applicable only to neutron energies above the Debye temperature of the sample. For solid \textit{ortho}-deuterium, the Debye temperature at low sample temperatures of $T=$ 0--18~K is around $\Theta_\text{D}=110$~K~\cite{hill:1959,nielsen:1973}, and for liquid \textit{ortho}-deuterium at $T=20$~K it is $\Theta_\text{D}=101$~K~\cite{guarini:2016,guarini:2015-privcomm}, equivalent to 9~meV, which is the energy of subthermal neutrons.

Previously, we published a parameter-free analytical calculation model for the scattering cross sections of liquid deuterium for slow neutrons~\cite{doege:2015}, especially in the UCN and the VCN energy range. The group of Guarini~\textit{et al.} was able to calculate the fundamental properties of the hydrogen liquids from parameter-free quantum simulations~\cite{guarini:2015,guarini:2016} and to derive the scattering cross sections of liquid \textit{ortho}-deuterium for thermal and cold neutrons. For a more detailed treatment of the scattering theory for liquid \textit{ortho}-deuterium, the reader is kindly referred to these three publications.

\section{Experiment Setup}\label{sec:exp-setup}

The neutron scattering cross section $\sigma_\text{scatt}$ is measured by means of a transmission experiment. To this end, we employed the standard transmission equation for uniformly absorbing and scattering media (in optics known as the Lambert--Beer law~\cite{bouguer:1729, beer:1852}),
\begin{equation}\label{eq:transmission-eq}
T(v) = \frac{I(d_n)}{I_0} = \text{e}^{-N_\text{v}\sigma_\text{tot}(v) d_n},
\end{equation}
where $T(v)$ represents the measured absolute transmissivity of the sample, $I_0$ the neutron beam intensity behind the empty sample container, $I(d_n)$ the neutron beam intensity behind the container filled with sample $n$, $N_\text{v}$ the particle number density of the sample, $d_n$ the sample thickness, and $\sigma_\text{tot}(v)$ the \textit{total} UCN loss cross section of the sample bulk for the neutron velocity $v$ in the medium, i.e., taking into account the neutron-optical potential of the sample. Equation~\ref{eq:transmission-eq} is solved for $\sigma_\text{tot}$, which is then corrected for the absorption of neutrons in the sample, sample impurities, and other side effects to yield the total \textit{scattering} cross section $\sigma_\text{scatt}=\sigma_\text{tot}-\sigma_\text{abs,D}-\sigma_\text{abs,H}$. It should be noted that, in the case of the liquid \textit{ortho}-deuterium used here, these corrections amounted to less than 1\% over the entire UCN range.

The equipment for this experiment was set up at the ``Turbine''~\cite{steyerl:1986} beamline PF2-EDM of the Institut Laue--Langevin (ILL) in Grenoble, France. Before the measurements were carried out, the neutron flux at that beamline was characterized~\cite{doege:2020-turbine}. This provided a solid foundation for the data treatment.

\begin{figure}[!h]
	\centering
	\includegraphics[width=1.00\columnwidth]{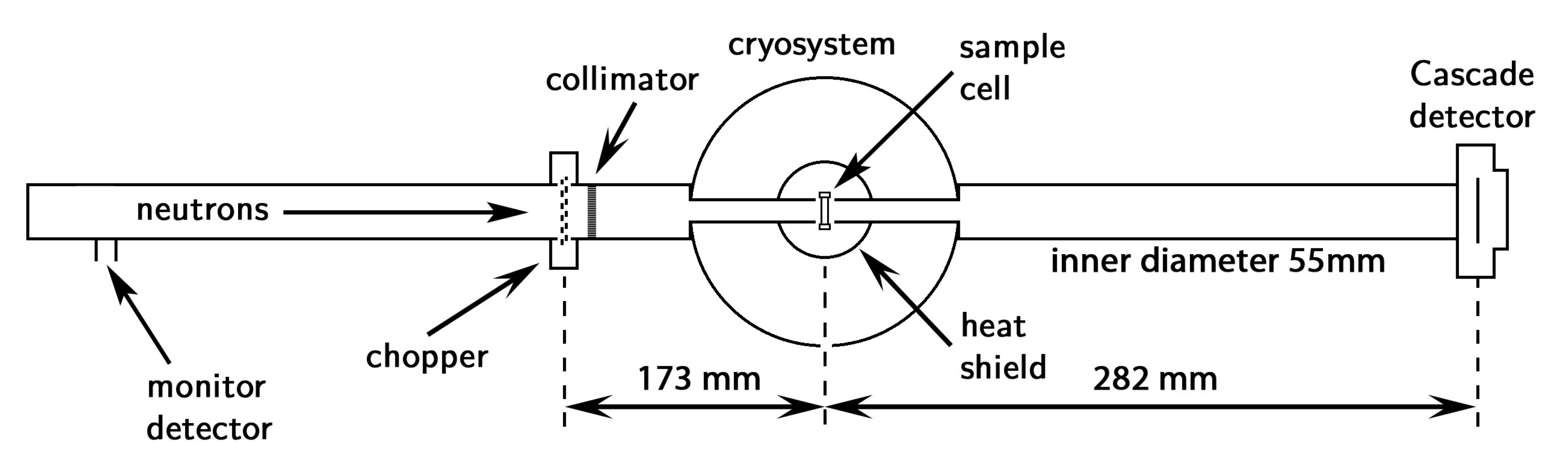}
  \caption[TOF geometry]{The TOF geometry of the UCN transmission measurements on liquid \textit{ortho}-deuterium.}\label{fig:tof-geometry-2016}
\end{figure}

The experiment itself was performed using a neutron time-of-flight (TOF) setup similar to that described earlier~\cite{doege:2015}. This time, the chopper had titanium shutters and was placed upstream of the liquid \textit{ortho}-deuterium sample, see Fig.~\ref{fig:tof-geometry-2016}. The chopper's opening function was verified in a measurement with light, where its full width at half maximum (FWHM) was determined to be $\tau=13.6$~ms. The total flight path was 455~mm long -- a compromise between having a maximum path length while not losing too many slow UCNs due to them falling in the gravitational field and being absorbed by or scattered on the flight tube's walls. A monitor detector enabled the tracking of possible changes of the UCN flux, which were less than 1\% over the course of the experiment.

The TOF geometry of the experiment demands that all neutrons travel the same path length. Therefore, they need to be collimated before they impinge on the sample. Neutrons diverging from the flight path need to be effectively removed from the experiment. This point was addressed by using a polyethylene foil as a liner in the flight path between the sample and the detector, which up-scatters UCNs to thermal energies, thus removing them from the measurable UCN spectrum. The collimator used here was a disk made from titanium instead of one made from poly(methyl methacrylate) (PMMA), as used earlier. UCN absorption in titanium leads to a significantly reduced background in the measurement as compared to UCN up-scattering in PMMA.

Furthermore, the body of the sample container was made from copper to reduce the temperature gradient across the sample volume to a minimum compared to the previously used aluminum sample container.

As we have shown earlier, rough surfaces cause tremendous scattering of UCNs~\cite{doege:2020-foils}. In transmission measurements, this can translate to significant experimental errors. Furthermore, previous experiments by various groups have shown that aluminum windows tend to bulge under pressure~\cite{atchison:2005-liq} leading to poorly defined sample thicknesses and introducing avoidable errors to the data treatment procedure. To reduce these potential error sources and to improve counting statistics, we developed and used in this experiment a sample container with low-roughness, transparent windows~\cite{doege:2018}.

In the flight tube between the cryostat and the neutron detector, a horizontal viewport was installed. In conjunction with a movable mirror on a rod and the transparent sample cell windows, it allowed observation of the sample preparation process in real time along the neutron flight path. When the sample was ready to be measured with neutrons, the mirror was raised out of the flight path.

All improvements over our 2012 experiment~\cite{doege:2015} led to a tenfold increased signal-to-noise ratio~\cite{doege:2019-phd}.

The deuterium gas in our experiment had a purity grade of N30, meaning at least 99.90\% deuterium. The temperature gradient across the sample container was $\Delta T = \pm 0.25$~K and the liquid samples were held at the vapor pressure of deuterium for their respective temperature during the entire measurement run. We used Raman spectroscopy as described by Silvera~\cite{silvera:1980} to determine the \textit{ortho}-deuterium fraction in the sample, which was $c_\text{ortho}=97.2\pm 0.02$\%.

For the registration of UCN counts~\cite{doege:2017-3-14-374}, a boron-lined gas electron multiplier (GEM) detector of the Cascade type~\cite{klein:2011} was used with a flushing and quenching gas mixture of 90\% argon and 10\% carbon dioxide.

\section{Cross-Section Results and Comparison with Model Calculations}\label{sec:exp-results}

\subsection{Deconvolution and data treatment}

As required by Eq.~\ref{eq:transmission-eq}, two UCN transmission spectra were recorded per measurement -- one with an empty sample container and one with liquid deuterium at the respective temperature. Both measured spectra are convolved with the chopper's resolution function. To account for this, the spectra needed to be deconvolved, for which a standard procedure is lacking. The deconvolution method presented here improves upon the data presented in Ref.~\cite{doege:2019-phd}.

The spectra were prepared for deconvolution by subtracting background counts, normalizing them to the measurement time, binning them using a sliding frame of 8~ms width, transforming the time into the velocity domain, and correcting for neutron reflection at interfaces.

The velocity-dependent transmission equation of the measured spectra $\widetilde{I}(d_n)$ and $\widetilde{I}_0$, i.e., convolved with the resolution function $G(v)$, can be expressed as
\begin{equation}\label{eq:transmission-eq-conv}
\frac{\widetilde{I}(d_n)}{\widetilde{I}_0} = \frac{I(d_n) \ast G(v)}{I_0 \ast G(v)} = \frac{[M(v) T^\text{expt}(v)]\ast G(v)}{M(v) \ast G(v)},
\end{equation}
\noindent
where $T^\text{expt}(v) = \text{e}^{-N_\text{v}\sigma_\text{tot} d_n}$ is the true neutron transmission from Eq.~\ref{eq:transmission-eq} with $\sigma_\text{tot} \propto 1/v$~\cite{lovesey:1984} in the liquid phase, $I_0 = M(v)$ is the known Maxwellian spectrum of the neutron source PF2-EDM~\cite{doege:2020-turbine}, and $G(v)$ is the chopper's resolution function represented by a normalized Gaussian function
\begin{equation}
G(v) = \frac{1}{\sqrt{2\pi} \sigma_\text{G}} \text{e}^{-\frac{1}{2}\left(\frac{v}{\sigma_\text{G}}\right)^2},
\end{equation}
with $\sigma_\text{G}$ the standard deviation of the Gaussian, which relates to its FWHM by $2\sqrt{2 \ln 2}\sigma_\text{G}$.

In our TOF experiment with neutrons, the neutron pulse had a length of $\tau = 13.6$~ms (FWHM) at the position of the chopper. The standard deviation $\sigma_\text{G}$ of the Gaussian is velocity-dependent due to the dispersion of the neutron pulse as it travels along the flight path $s_0 = 0.455$~m,
\begin{equation}\label{eq:velocity-dep-opening}
\sigma_\text{G}(v) = \frac{1}{\sqrt{8 \ln 2}} \frac{v^2}{s_0/\tau}.
\end{equation}

To calculate the deconvolution factors $D(v)$ for our experimental datasets $\widetilde{I}_0$ and $\widetilde{I}(d_n)$, see Eq.~\ref{eq:transmission-eq-conv}, we first convolved both the known unperturbed neutron spectrum $I_0 = M(v)$ of the neutron source and the calculated neutron spectrum behind the sample, $I(d_n) = I_0 T^\text{th}(v)$, with $G(v)$, the Gaussian resolution function. The scattering cross section $\sigma_\text{tot}=\sigma_\text{scatt}$ for calculating $T^\text{th}(v)$ was taken from the calculation model~\cite{doege:2015}, see Fig.~\ref{fig:cross-section-liquid-deuterium}. 

The convolution factors $C(v)$ reproduce the effect of the convolution on both spectra $I_0$ and $I(d_n)$, and depend on the neutron velocity. They are defined as
\begin{equation}
\frac{[M(v) T^\text{th}(v)]\ast G(v)}{M(v) \ast G(v)} \stackrel{\text{def}}{=} C(v) \frac{[M(v) T^\text{th}(v)]}{M(v)} = C(v) T^\text{th}(v).
\end{equation}
Since the convolution has a slightly different effect on $I_0$ and $I(d_n) = I_0 T^\text{th}(v)$, the transmission $T^\text{th}$ is changed by the convolution as well.

Now we can apply the inverse of the convolution factors $C(v)$, namely the deconvolution factor $D(v) = C^{-1}(v)$, to deconvolve the \textit{measured} spectra and retrieve the corrected transmission,
\begin{equation}
D(v) \frac{\widetilde{I}(d_n)}{\widetilde{I}_0} = T(v).
\end{equation}

$D(v)$ is a smooth function with $D = 1$ around $v=8$~m/s, the maximum of the incoming neutron spectrum. The corrected transmission $T(v)$ enters into Eq.~\ref{eq:transmission-eq}, from which $\sigma_\text{tot}$ is calculated.

In the case of our experiment, the total cross sections of neutrons with a velocity below 9.5~m/s are virtually unaffected by the deconvolution, changing them only by 1\% or less. For neutron velocities between 10 and 15~m/s, the deconvolution suppresses $\sigma_\text{tot}$ from 2\% to 15\%, respectively, due to the large velocity uncertainty as shown by Eq.~\ref{eq:velocity-dep-opening}.

After deconvolution and calculation of $\sigma_\text{tot}$, the results were corrected for absorption by deuterium ($\sigma_\text{abs,D} = 2.28/v [\text{b} \times \text{m/s}]$)~\cite{sears:1992} and hydrogen impurities ($\sigma_\text{abs,H}$) in the sample to obtain $\sigma_\text{scatt}$. Details are described in Ref.~\cite{doege:2019-phd}.

\subsection{Experimental results of scattering cross-section measurements}

\begin{figure}[!ht]
\centering
\includegraphics[width=0.95\columnwidth]{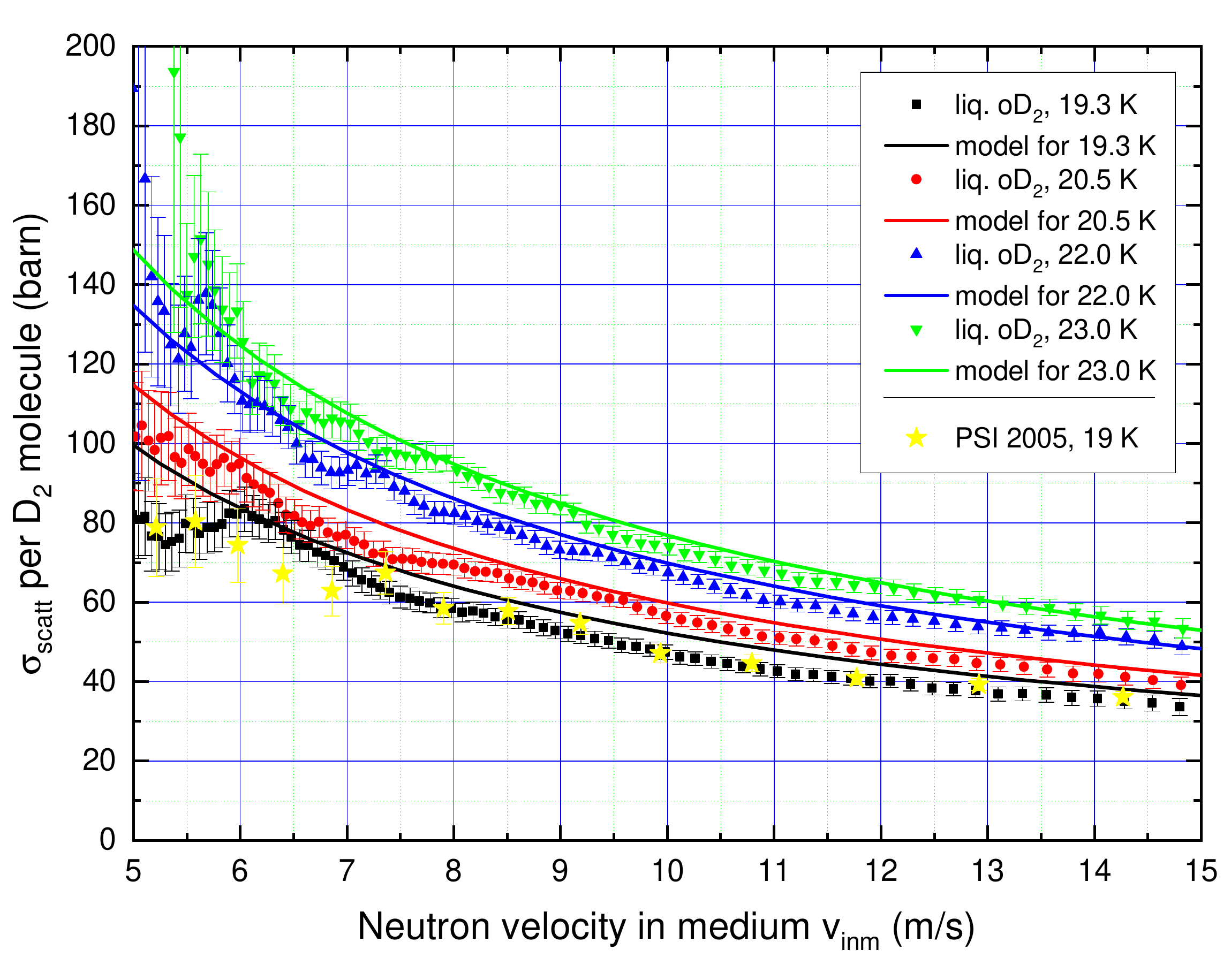}
\caption[Scattering cross section of liquid deuterium]{\label{fig:cross-section-liquid-deuterium}Experimental data points (solid colored symbols) for the scattering cross section of liquid \textit{ortho}-deuterium at various temperatures. The theoretical model by D\"{o}ge \textit{et al.}~\cite{doege:2015} was calculated for the same temperatures as in this experiment using $c_\text{ortho} = 0.972$ (solid colored lines) and the self-diffusion coefficient for liquid deuterium from O'Reilly and Peterson~\cite{oreilly:1977}. The temperature uncertainty in the experiment was $\Delta T= \pm 0.25$~K. The \textit{total} cross section data from Atchison \textit{et al.}~\cite{atchison:2005-liq} are shown for comparison (PSI 2005, yellow stars) and match both our measured data and the calculation model well. Our scattering cross sections were corrected for absorption on H$_2$ and D$_2$, and the Atchison \textit{et al.} data were not. This correction falls within the error bars and is thus negligible.}
\end{figure}

The UCN scattering cross sections $\sigma_\text{scatt}$ of liquid \textit{ortho}-deuterium ($c_\text{ortho} = 0.972$) for the temperatures 19.3~K, 20.5~K, 22.0~K, and 23.0~K, corrected for absorption in the medium and for reflection at the sample interfaces, are shown in Fig.~\ref{fig:cross-section-liquid-deuterium}. They include quasi-elastic contributions arising from the heat conductivity and particle diffusion, as well as one-phonon up-scattering in \textit{ortho}-deuterium ($\sigma_\text{00}$). Contributions from \textit{para}-deuterium impurities, which play a negligible role due to the low \textit{para}-D$_2$ concentration, are quasi-elastic and inelastic scattering ($\sigma_\text{11}$) as well as up-scattering due to the rotational relaxation of \textit{para}-deuterium ($\sigma_\text{10}$). The scattering cross sections obtained from the measurement and after corrections are composed of the following constituents
\begin{equation}
\sigma_\text{scatt} (v) = c_\text{ortho}\times \sigma_\text{00} (v) + c_\text{para} \times \left[\sigma_\text{11} (v) + \sigma_\text{10} (v) \right],
\end{equation}
where $c_\text{ortho}$ and $c_\text{para}$ are the \textit{ortho}- and \textit{para}-deuterium concentrations in the sample and $\sigma_{JJ'}$ the deuterium molecular scattering cross sections for the respective rotational states of the deuterium molecule before ($J$) and after scattering ($J'$)~\cite{liu:2000,doege:2015}.

\subsection{Comparison with model calculations and previous experimental results}

The experimental data presented in Fig.~\ref{fig:cross-section-liquid-deuterium} agree very well with the values obtained for the same four temperatures using for the double-differential scattering cross section $\text{d}^2\sigma/\text{d}\Omega \text{d}E$ the calculation model published by D\"{o}ge et al.~\cite{doege:2015}. It calculates the coherent and incoherent scattering contributions in full according to
\begin{equation}\label{eq:d-diff_coh-inc}
\frac{\text{d}^2 \sigma}{\text{d}\Omega \text{d}E} = \frac{1}{4\pi}\frac{k_\text{f}}{k_\text{i}} \left[\sigma^\text{coh} \times S_\text{coh}(q,\omega) + \sigma^\text{inc} \times S_\text{inc}(q,\omega) \right],
\end{equation}
\noindent
where $k_\text{i}$ is the wave vector of the incoming neutron, $k_\text{f}$ that of the scattered neutron, $\sigma^\text{coh,inc}$ are the coherent and incoherent molecular scattering cross sections, and $S_\text{coh,inc}(q,\omega)$ are the coherent and incoherent scattering laws, respectively.

Both experimental and calculated scattering cross sections are inversely proportional to the neutron's velocity.

It must be noted that instead of the self-diffusion coefficient based on unpublished data, which was used in Ref.~\cite{doege:2015}, now the experimentally determined values from O'Reilly and Peterson~\cite{oreilly:1977} were used in this model. Quantum centroid molecular dynamics simulations by Guarini \textit{et al.}~\cite{guarini:2016} confirm these values, e.g., $D_\text{s}=3.5\times 10^{-5}$~cm$^2$/s~\cite{guarini:2015-privcomm} and $D_\text{s}=(3.7\pm 0.4)\times 10^{-5}$~cm$^2$/s~\cite{oreilly:1977}, both for $T=20$~K. The values of the other thermophysical properties of deuterium, upon which the calculation model relies, were taken from Souers~\cite{souers:1986}.

These experimental data on liquid deuterium published here should be considered an improvement of the experimental data published in Refs.~\cite{doege:2015,doege:2016}, which showed a large temperature gradient across the sample container and were, in retrospect, probably taken at a higher temperature than the average between the two measurement points in the sample container.

Our scattering cross sections shown in Fig.~\ref{fig:cross-section-liquid-deuterium} were corrected for absorption on H$_2$ and D$_2$, and the Atchison \textit{et al.} data (PSI 2005) were not. This correction falls within the error bars and does not impair the comparability of both datasets.

\section{Discussion}

Considering the high degree of agreement between the experimentally measured scattering cross sections and the calculated ones at four different temperatures covering the entire liquid phase of deuterium, we can confidently say that our calculation model proved accurate. The agreement between our data for 19.3~K (taken in a sample container with low-roughness windows) and the data from Atchison \textit{et al.} for 19~K (taken in a sample container with machined aluminum windows that had a high surface roughness) implies that sample containers with rough surfaces cause only negligible errors in the measurement of liquid samples. This is in stark contrast with their adverse effect on solid samples~\cite{doege:2019-phd}.

In our previous paper on liquid deuterium~\cite{doege:2015}, we concluded that the incoherent approximation~\cite{turchin:1965} was valid for ultracold-neutron scattering in liquid deuterium. After recalculating the scattering model from Ref.~\cite{doege:2015} for both coherent and incoherent scattering contributions, see Eq.~\ref{eq:d-diff_coh-inc}, using the diffusion coefficient from O'Reilly and Peterson, we cannot uphold this conclusion. In fact, the incoherent approximation overestimates the scattering cross sections of \textit{ortho}-deuterium by a factor of 3 and that of \textit{para}-deuterium by a factor of 4 over the entire temperature range of the liquid phase. These factors are comparable to the overestimation that the incoherent approximation delivers for ultracold-neutron scattering in solid \textit{ortho}-deuterium~\cite{doege:2021-inc-approx}.

\section{Conclusion}

We have provided new experimental scattering cross section data of liquid \textit{ortho}-deuterium for ultracold neutrons (UCNs) at four different temperatures. Our results for 19.3~K are very similar to those obtained by Atchison \textit{et al.} at 19~K~\cite{atchison:2005-liq}. This implies that sample containers with rough surfaces cause only negligible errors when measuring liquid samples, whereas they have detrimental effects on solid samples.

Furthermore, our experimental data show a high degree of agreement with -- and therefore confirm -- a calculation model published by D\"{o}ge \textit{et al.}~\cite{doege:2015}, which has no free parameters and uses only the material properties of deuterium. One key variable in this calculation model is the self-diffusion constant of liquid deuterium, for which we took the measured value from O'Reilly and Peterson~\cite{oreilly:1977}. This value was confirmed by quantum centroid molecular dynamics simulations by Guarini \textit{et al.}~\cite{guarini:2016}.

Both experimental and calculated scattering cross sections are inversely proportional to the neutron's velocity.

We conclude that the calculation model of D\"{o}ge \textit{et al.} is suitable for calculating the scattering cross sections of liquid deuterium in the ultracold and very cold-neutron ranges for arbitrary \textit{ortho} and \textit{para} concentrations, and for all temperatures of the liquid phase of deuterium. It can, therefore, be used to model scattering kernels S($\alpha$,$\beta$) closer to the underlying physics of liquid deuterium and to extend existing kernels, e.g. from Bernnat~\cite{bernnat:2002}, to the low-energy UCN range. In addition, we have shown that, applying scattering kernels, the incoherent approximation cannot be used to calculate the coherent scattering cross sections in the UCN range.

The oft applied simplification called the incoherent approximation, which neglects interference effects in the neutron-scattering process, overestimates the ultracold-neutron scattering cross sections of liquid deuterium by a factor of 3--4. This is in line with our previous calculations and experiments on solid deuterium~\cite{doege:2019-phd,doege:2021-inc-approx}.

The results for this paper were produced as part of the Ph.D. thesis of Stefan D\"oge~\cite{doege:2019-phd}.

\begin{acknowledgments}
We are grateful to Eleonora Guarini of Universit\`a degli Studi di Firenze for fruitful discussions and for providing the self-diffusion coefficient of liquid deuterium from quantum simulations. Furthermore, we wish to thank the reactor crew, the instrument scientists, and the technicians of the mechanical workshops at the Institut Laue--Langevin, Grenoble, for their support during the beamtime no. 3-14-374. We gratefully acknowledge Xavier Tonon and Eddy Leli\`evre-Berna of the ILL Cryogenics Service for valuable advice in improving the cryostat, and Brian C. Dye from Dallas, TX, USA for a critical reading of the manuscript. Bernhard Lauss from the UCN group of the Paul Scherrer Institut (PSI) in Villigen, Switzerland provided a Raman spectrometer for our experiments and Nicolas Hild (also from the PSI) evaluated the Raman measurements. This work received funding from FRM~II/ Heinz Maier-Leibnitz Zentrum (MLZ), Munich, Germany. The open access publication fee was paid for by the University Library and the Physics Department of the Technische Universit\"at M\"unchen, Munich, Germany.


\end{acknowledgments}


\begin{thebibliography}{44}%
\makeatletter
\providecommand \@ifxundefined [1]{%
 \@ifx{#1\undefined}
}%
\providecommand \@ifnum [1]{%
 \ifnum #1\expandafter \@firstoftwo
 \else \expandafter \@secondoftwo
 \fi
}%
\providecommand \@ifx [1]{%
 \ifx #1\expandafter \@firstoftwo
 \else \expandafter \@secondoftwo
 \fi
}%
\providecommand \natexlab [1]{#1}%
\providecommand \enquote  [1]{``#1''}%
\providecommand \bibnamefont  [1]{#1}%
\providecommand \bibfnamefont [1]{#1}%
\providecommand \citenamefont [1]{#1}%
\providecommand \href@noop [0]{\@secondoftwo}%
\providecommand \href [0]{\begingroup \@sanitize@url \@href}%
\providecommand \@href[1]{\@@startlink{#1}\@@href}%
\providecommand \@@href[1]{\endgroup#1\@@endlink}%
\providecommand \@sanitize@url [0]{\catcode `\\12\catcode `\$12\catcode
  `\&12\catcode `\#12\catcode `\^12\catcode `\_12\catcode `\%12\relax}%
\providecommand \@@startlink[1]{}%
\providecommand \@@endlink[0]{}%
\providecommand \url  [0]{\begingroup\@sanitize@url \@url }%
\providecommand \@url [1]{\endgroup\@href {#1}{\urlprefix }}%
\providecommand \urlprefix  [0]{URL }%
\providecommand \Eprint [0]{\href }%
\providecommand \doibase [0]{https://doi.org/}%
\providecommand \selectlanguage [0]{\@gobble}%
\providecommand \bibinfo  [0]{\@secondoftwo}%
\providecommand \bibfield  [0]{\@secondoftwo}%
\providecommand \translation [1]{[#1]}%
\providecommand \BibitemOpen [0]{}%
\providecommand \bibitemStop [0]{}%
\providecommand \bibitemNoStop [0]{.\EOS\space}%
\providecommand \EOS [0]{\spacefactor3000\relax}%
\providecommand \BibitemShut  [1]{\csname bibitem#1\endcsname}%
\let\auto@bib@innerbib\@empty
\bibitem [{\citenamefont {Eucken}(1912)}]{eucken:1912}%
  \BibitemOpen
  \bibfield  {author} {\bibinfo {author} {\bibfnamefont {A.}~\bibnamefont
  {Eucken}},\ }\bibfield  {title} {\bibinfo {title} {{D}ie {M}olekularw\"{a}rme
  des {W}asserstoffs bei tiefen {T}emperaturen},\ }\href@noop {} {\bibfield
  {journal} {\bibinfo  {journal} {Sitzber. Preuss. Akad. Wiss.}\ ,\ \bibinfo
  {pages} {141}} (\bibinfo {year} {1912})}\BibitemShut {NoStop}%
\bibitem [{\citenamefont {Dennison}(1927)}]{dennison:1927}%
  \BibitemOpen
  \bibfield  {author} {\bibinfo {author} {\bibfnamefont {D.~M.}\ \bibnamefont
  {Dennison}},\ }\bibfield  {title} {\bibinfo {title} {A note on the specific
  heat of the hydrogen molecule},\ }\href {http://www.jstor.org/stable/94849}
  {\bibfield  {journal} {\bibinfo  {journal} {Proceedings of the Royal Society
  of London. Series A, Containing Papers of a Mathematical and Physical
  Character}\ }\textbf {\bibinfo {volume} {115}},\ \bibinfo {pages} {483}
  (\bibinfo {year} {1927})}\BibitemShut {NoStop}%
\bibitem [{\citenamefont {Heisenberg}(1927)}]{heisenberg:1927}%
  \BibitemOpen
  \bibfield  {author} {\bibinfo {author} {\bibfnamefont {W.}~\bibnamefont
  {Heisenberg}},\ }\bibfield  {title} {\bibinfo {title}
  {{M}ehrk{\"o}rperprobleme und {R}esonanz in der {Q}uantenmechanik. {II}},\
  }\href {https://doi.org/10.1007/BF01391241} {\bibfield  {journal} {\bibinfo
  {journal} {Zeitschrift f{\"u}r Physik A Hadrons and Nuclei}\ }\textbf
  {\bibinfo {volume} {41}},\ \bibinfo {pages} {239} (\bibinfo {year}
  {1927})}\BibitemShut {NoStop}%
\bibitem [{\citenamefont {Hund}(1927)}]{hund:1927}%
  \BibitemOpen
  \bibfield  {author} {\bibinfo {author} {\bibfnamefont {F.}~\bibnamefont
  {Hund}},\ }\bibfield  {title} {\bibinfo {title} {Zur {D}eutung der
  {M}olekelspektren. {II}},\ }\href {https://doi.org/10.1007/BF01397124}
  {\bibfield  {journal} {\bibinfo  {journal} {Zeitschrift f{\"u}r Physik}\
  }\textbf {\bibinfo {volume} {42}},\ \bibinfo {pages} {93} (\bibinfo {year}
  {1927})}\BibitemShut {NoStop}%
\bibitem [{\citenamefont {Farkas}\ \emph {et~al.}(1934)\citenamefont {Farkas},
  \citenamefont {Farkas}, \citenamefont {Harteck},\ and\ \citenamefont
  {Rideal}}]{farkas:1934}%
  \BibitemOpen
  \bibfield  {author} {\bibinfo {author} {\bibfnamefont {A.}~\bibnamefont
  {Farkas}}, \bibinfo {author} {\bibfnamefont {L.}~\bibnamefont {Farkas}},
  \bibinfo {author} {\bibfnamefont {P.}~\bibnamefont {Harteck}},\ and\ \bibinfo
  {author} {\bibfnamefont {E.~K.}\ \bibnamefont {Rideal}},\ }\bibfield  {title}
  {\bibinfo {title} {{E}xperiments on heavy hydrogen. {II}. {T}he ortho-para
  conversion},\ }\href {https://doi.org/10.1098/rspa.1934.0062} {\bibfield
  {journal} {\bibinfo  {journal} {Proceedings of the Royal Society of London.
  Series A, Containing Papers of a Mathematical and Physical Character}\
  }\textbf {\bibinfo {volume} {144}},\ \bibinfo {pages} {481} (\bibinfo {year}
  {1934})}\BibitemShut {NoStop}%
\bibitem [{\citenamefont {Silvera}(1980)}]{silvera:1980}%
  \BibitemOpen
  \bibfield  {author} {\bibinfo {author} {\bibfnamefont {I.~F.}\ \bibnamefont
  {Silvera}},\ }\bibfield  {title} {\bibinfo {title} {{T}he solid molecular
  hydrogens in the condensed phase: {F}undamentals and static properties},\
  }\href {https://doi.org/10.1103/RevModPhys.52.393} {\bibfield  {journal}
  {\bibinfo  {journal} {Rev. Mod. Phys.}\ }\textbf {\bibinfo {volume} {52}},\
  \bibinfo {pages} {393} (\bibinfo {year} {1980})}\BibitemShut {NoStop}%
\bibitem [{\citenamefont {Ageron}\ \emph {et~al.}(1969)\citenamefont {Ageron},
  \citenamefont {{De Beaucourt}}, \citenamefont {Harig}, \citenamefont
  {Lacaze},\ and\ \citenamefont {Livolant}}]{ageron:1969}%
  \BibitemOpen
  \bibfield  {author} {\bibinfo {author} {\bibfnamefont {P.}~\bibnamefont
  {Ageron}}, \bibinfo {author} {\bibfnamefont {P.}~\bibnamefont {{De
  Beaucourt}}}, \bibinfo {author} {\bibfnamefont {H.}~\bibnamefont {Harig}},
  \bibinfo {author} {\bibfnamefont {A.}~\bibnamefont {Lacaze}},\ and\ \bibinfo
  {author} {\bibfnamefont {M.}~\bibnamefont {Livolant}},\ }\bibfield  {title}
  {\bibinfo {title} {Experimental and theoretical study of cold neutron sources
  of liquid hydrogen and liquid deuterium},\ }\href
  {https://doi.org/https://doi.org/10.1016/0011-2275(69)90257-4} {\bibfield
  {journal} {\bibinfo  {journal} {Cryogenics}\ }\textbf {\bibinfo {volume}
  {9}},\ \bibinfo {pages} {42} (\bibinfo {year} {1969})}\BibitemShut {NoStop}%
\bibitem [{\citenamefont {Golub}\ \emph {et~al.}(1991)\citenamefont {Golub},
  \citenamefont {Richardson},\ and\ \citenamefont {Lamoreaux}}]{golub:1991}%
  \BibitemOpen
  \bibfield  {author} {\bibinfo {author} {\bibfnamefont {R.}~\bibnamefont
  {Golub}}, \bibinfo {author} {\bibfnamefont {D.~J.}\ \bibnamefont
  {Richardson}},\ and\ \bibinfo {author} {\bibfnamefont {S.~K.}\ \bibnamefont
  {Lamoreaux}},\ }\href@noop {} {\emph {\bibinfo {title} {Ultra-Cold
  Neutrons}}}\ (\bibinfo  {publisher} {Adam Hilger, Bristol, U.K.},\ \bibinfo
  {year} {1991})\BibitemShut {NoStop}%
\bibitem [{\citenamefont {Lychagin}\ \emph {et~al.}(2016)\citenamefont
  {Lychagin}, \citenamefont {Mityukhlyaev}, \citenamefont {Muzychka},
  \citenamefont {Nekhaev}, \citenamefont {Nesvizhevsky}, \citenamefont
  {Onegin}, \citenamefont {Sharapov},\ and\ \citenamefont
  {Strelkov}}]{lychagin:2016}%
  \BibitemOpen
  \bibfield  {author} {\bibinfo {author} {\bibfnamefont {E.}~\bibnamefont
  {Lychagin}}, \bibinfo {author} {\bibfnamefont {V.}~\bibnamefont
  {Mityukhlyaev}}, \bibinfo {author} {\bibfnamefont {A.}~\bibnamefont
  {Muzychka}}, \bibinfo {author} {\bibfnamefont {G.}~\bibnamefont {Nekhaev}},
  \bibinfo {author} {\bibfnamefont {V.}~\bibnamefont {Nesvizhevsky}}, \bibinfo
  {author} {\bibfnamefont {M.}~\bibnamefont {Onegin}}, \bibinfo {author}
  {\bibfnamefont {E.}~\bibnamefont {Sharapov}},\ and\ \bibinfo {author}
  {\bibfnamefont {A.}~\bibnamefont {Strelkov}},\ }\bibfield  {title} {\bibinfo
  {title} {{UCN} sources at external beams of thermal neutrons. {A}n example of
  {PIK} reactor},\ }\href {https://doi.org/10.1016/j.nima.2016.04.008}
  {\bibfield  {journal} {\bibinfo  {journal} {Nuclear Instruments and Methods
  in Physics Research Section A: Accelerators, Spectrometers, Detectors and
  Associated Equipment}\ }\textbf {\bibinfo {volume} {823}},\ \bibinfo {pages}
  {47} (\bibinfo {year} {2016})}\BibitemShut {NoStop}%
\bibitem [{\citenamefont {Schreyer}\ \emph {et~al.}(2020)\citenamefont
  {Schreyer}, \citenamefont {Davis}, \citenamefont {Kawasaki}, \citenamefont
  {Kikawa}, \citenamefont {Marshall}, \citenamefont {Mishima}, \citenamefont
  {Okamura},\ and\ \citenamefont {Picker}}]{schreyer:2020}%
  \BibitemOpen
  \bibfield  {author} {\bibinfo {author} {\bibfnamefont {W.}~\bibnamefont
  {Schreyer}}, \bibinfo {author} {\bibfnamefont {C.}~\bibnamefont {Davis}},
  \bibinfo {author} {\bibfnamefont {S.}~\bibnamefont {Kawasaki}}, \bibinfo
  {author} {\bibfnamefont {T.}~\bibnamefont {Kikawa}}, \bibinfo {author}
  {\bibfnamefont {C.}~\bibnamefont {Marshall}}, \bibinfo {author}
  {\bibfnamefont {K.}~\bibnamefont {Mishima}}, \bibinfo {author} {\bibfnamefont
  {T.}~\bibnamefont {Okamura}},\ and\ \bibinfo {author} {\bibfnamefont
  {R.}~\bibnamefont {Picker}},\ }\bibfield  {title} {\bibinfo {title}
  {Optimizing neutron moderators for a spallation-driven ultracold-neutron
  source at {TRIUMF}},\ }\href {https://doi.org/10.1016/j.nima.2020.163525}
  {\bibfield  {journal} {\bibinfo  {journal} {Nuclear Instruments and Methods
  in Physics Research Section A: Accelerators, Spectrometers, Detectors and
  Associated Equipment}\ }\textbf {\bibinfo {volume} {959}},\ \bibinfo {pages}
  {163525} (\bibinfo {year} {2020})}\BibitemShut {NoStop}%
\bibitem [{\citenamefont {Dubbers}\ and\ \citenamefont
  {Schmidt}(2011)}]{dubbers:2011}%
  \BibitemOpen
  \bibfield  {author} {\bibinfo {author} {\bibfnamefont {D.}~\bibnamefont
  {Dubbers}}\ and\ \bibinfo {author} {\bibfnamefont {M.~G.}\ \bibnamefont
  {Schmidt}},\ }\bibfield  {title} {\bibinfo {title} {The neutron and its role
  in cosmology and particle physics},\ }\href
  {https://doi.org/10.1103/RevModPhys.83.1111} {\bibfield  {journal} {\bibinfo
  {journal} {Rev. Mod. Phys.}\ }\textbf {\bibinfo {volume} {83}},\ \bibinfo
  {pages} {1111} (\bibinfo {year} {2011})}\BibitemShut {NoStop}%
\bibitem [{\citenamefont {Crivellin}\ and\ \citenamefont
  {Hoferichter}(2020)}]{crivellin:2020}%
  \BibitemOpen
  \bibfield  {author} {\bibinfo {author} {\bibfnamefont {A.}~\bibnamefont
  {Crivellin}}\ and\ \bibinfo {author} {\bibfnamefont {M.}~\bibnamefont
  {Hoferichter}},\ }\bibfield  {title} {\bibinfo {title} {$\ensuremath{\beta}$
  {D}ecays as sensitive probes of lepton flavor universality},\ }\href
  {https://doi.org/10.1103/PhysRevLett.125.111801} {\bibfield  {journal}
  {\bibinfo  {journal} {Phys. Rev. Lett.}\ }\textbf {\bibinfo {volume} {125}},\
  \bibinfo {pages} {111801} (\bibinfo {year} {2020})}\BibitemShut {NoStop}%
\bibitem [{\citenamefont {Seiffert}(1970)}]{seiffert:1970}%
  \BibitemOpen
  \bibfield  {author} {\bibinfo {author} {\bibfnamefont {W.-D.}\ \bibnamefont
  {Seiffert}},\ }\emph {\bibinfo {title} {{M}essung der {S}treuquerschnitte von
  fl\"ussigem und festem {W}asserstoff, {D}euterium und {D}euteriumhydrid f\"ur
  thermische {N}eutronen}},\ \href@noop {} {Ph.D. thesis},\ \bibinfo  {school}
  {Technische Universit\"at M\"unchen, Munich, Germany} (\bibinfo {year}
  {1970}),\ \bibinfo {note} {also Euratom report no. EUR 4455 d}\BibitemShut
  {NoStop}%
\bibitem [{\citenamefont {Seiffert}\ \emph {et~al.}(1970)\citenamefont
  {Seiffert}, \citenamefont {Weckermann},\ and\ \citenamefont
  {Misenta}}]{seiffert:1970-article}%
  \BibitemOpen
  \bibfield  {author} {\bibinfo {author} {\bibfnamefont {W.~D.}\ \bibnamefont
  {Seiffert}}, \bibinfo {author} {\bibfnamefont {B.}~\bibnamefont
  {Weckermann}},\ and\ \bibinfo {author} {\bibfnamefont {R.}~\bibnamefont
  {Misenta}},\ }\bibfield  {title} {\bibinfo {title} {{M}essung der
  {S}treuquerschnitte von fl\"ussigem und festem {W}asserstoff, {D}euterium und
  {D}euteriumhydrid f\"ur thermische {N}eutronen},\ }\href
  {https://doi.org/10.1515/zna-1970-0626} {\bibfield  {journal} {\bibinfo
  {journal} {Zeitschrift für Naturforschung A}\ }\textbf {\bibinfo {volume}
  {25}},\ \bibinfo {pages} {967} (\bibinfo {year} {1970})}\BibitemShut
  {NoStop}%
\bibitem [{\citenamefont {Atchison}\ \emph {et~al.}(2005)\citenamefont
  {Atchison}, \citenamefont {van~den Brandt}, \citenamefont {Bry\'{s}},
  \citenamefont {Daum}, \citenamefont {Fierlinger}, \citenamefont {Hautle},
  \citenamefont {Henneck}, \citenamefont {Kirch}, \citenamefont {Kohlbrecher},
  \citenamefont {K\"uhne}, \citenamefont {Konter}, \citenamefont {Pichlmaier},
  \citenamefont {Wokaun}, \citenamefont {Bodek}, \citenamefont {Kasprzak},
  \citenamefont {Ku\'{z}niak}, \citenamefont {Geltenbort}, \citenamefont
  {Giersch}, \citenamefont {Zmeskal}, \citenamefont {Hino},\ and\ \citenamefont
  {Utsuro}}]{atchison:2005-liq}%
  \BibitemOpen
  \bibfield  {author} {\bibinfo {author} {\bibfnamefont {F.}~\bibnamefont
  {Atchison}}, \bibinfo {author} {\bibfnamefont {B.}~\bibnamefont {van~den
  Brandt}}, \bibinfo {author} {\bibfnamefont {T.}~\bibnamefont {Bry\'{s}}},
  \bibinfo {author} {\bibfnamefont {M.}~\bibnamefont {Daum}}, \bibinfo {author}
  {\bibfnamefont {P.}~\bibnamefont {Fierlinger}}, \bibinfo {author}
  {\bibfnamefont {P.}~\bibnamefont {Hautle}}, \bibinfo {author} {\bibfnamefont
  {R.}~\bibnamefont {Henneck}}, \bibinfo {author} {\bibfnamefont
  {K.}~\bibnamefont {Kirch}}, \bibinfo {author} {\bibfnamefont
  {J.}~\bibnamefont {Kohlbrecher}}, \bibinfo {author} {\bibfnamefont
  {G.}~\bibnamefont {K\"uhne}}, \bibinfo {author} {\bibfnamefont {J.~A.}\
  \bibnamefont {Konter}}, \bibinfo {author} {\bibfnamefont {A.}~\bibnamefont
  {Pichlmaier}}, \bibinfo {author} {\bibfnamefont {A.}~\bibnamefont {Wokaun}},
  \bibinfo {author} {\bibfnamefont {K.}~\bibnamefont {Bodek}}, \bibinfo
  {author} {\bibfnamefont {M.}~\bibnamefont {Kasprzak}}, \bibinfo {author}
  {\bibfnamefont {M.}~\bibnamefont {Ku\'{z}niak}}, \bibinfo {author}
  {\bibfnamefont {P.}~\bibnamefont {Geltenbort}}, \bibinfo {author}
  {\bibfnamefont {M.}~\bibnamefont {Giersch}}, \bibinfo {author} {\bibfnamefont
  {J.}~\bibnamefont {Zmeskal}}, \bibinfo {author} {\bibfnamefont
  {M.}~\bibnamefont {Hino}},\ and\ \bibinfo {author} {\bibfnamefont
  {M.}~\bibnamefont {Utsuro}},\ }\bibfield  {title} {\bibinfo {title} {Measured
  total cross sections of slow neutrons scattered by gaseous and liquid
  $^\text{2}${H}$_\text{2}$},\ }\href
  {https://doi.org/10.1103/PhysRevLett.94.212502} {\bibfield  {journal}
  {\bibinfo  {journal} {Phys. Rev. Lett.}\ }\textbf {\bibinfo {volume} {94}},\
  \bibinfo {pages} {212502} (\bibinfo {year} {2005})}\BibitemShut {NoStop}%
\bibitem [{\citenamefont {Squires}(1978)}]{squires:1978}%
  \BibitemOpen
  \bibfield  {author} {\bibinfo {author} {\bibfnamefont {G.~L.}\ \bibnamefont
  {Squires}},\ }\href@noop {} {\emph {\bibinfo {title} {Introduction to the
  Theory of Thermal Neutron Scattering}}}\ (\bibinfo  {publisher} {Dover
  Publications, Mineola, New York},\ \bibinfo {year} {1978})\BibitemShut
  {NoStop}%
\bibitem [{\citenamefont {Lovesey}(1984)}]{lovesey:1984}%
  \BibitemOpen
  \bibfield  {author} {\bibinfo {author} {\bibfnamefont {S.~W.}\ \bibnamefont
  {Lovesey}},\ }\href@noop {} {\emph {\bibinfo {title} {Theory of Neutron
  Scattering from Condensed Matter}}},\ Vol.~\bibinfo {volume} {1}\ (\bibinfo
  {publisher} {Clarendon Press, Oxford, U.K.},\ \bibinfo {year}
  {1984})\BibitemShut {NoStop}%
\bibitem [{\citenamefont {D\"oge}\ \emph {et~al.}(2015)\citenamefont {D\"oge},
  \citenamefont {Herold}, \citenamefont {M\"uller}, \citenamefont {Morkel},
  \citenamefont {Gutsmiedl}, \citenamefont {Geltenbort}, \citenamefont {Lauer},
  \citenamefont {Fierlinger}, \citenamefont {Petry},\ and\ \citenamefont
  {B\"oni}}]{doege:2015}%
  \BibitemOpen
  \bibfield  {author} {\bibinfo {author} {\bibfnamefont {S.}~\bibnamefont
  {D\"oge}}, \bibinfo {author} {\bibfnamefont {C.}~\bibnamefont {Herold}},
  \bibinfo {author} {\bibfnamefont {S.}~\bibnamefont {M\"uller}}, \bibinfo
  {author} {\bibfnamefont {C.}~\bibnamefont {Morkel}}, \bibinfo {author}
  {\bibfnamefont {E.}~\bibnamefont {Gutsmiedl}}, \bibinfo {author}
  {\bibfnamefont {P.}~\bibnamefont {Geltenbort}}, \bibinfo {author}
  {\bibfnamefont {T.}~\bibnamefont {Lauer}}, \bibinfo {author} {\bibfnamefont
  {P.}~\bibnamefont {Fierlinger}}, \bibinfo {author} {\bibfnamefont
  {W.}~\bibnamefont {Petry}},\ and\ \bibinfo {author} {\bibfnamefont
  {P.}~\bibnamefont {B\"oni}},\ }\bibfield  {title} {\bibinfo {title}
  {Scattering cross sections of liquid deuterium for ultracold neutrons:
  Experimental results and a calculation model},\ }\href
  {https://doi.org/10.1103/PhysRevB.91.214309} {\bibfield  {journal} {\bibinfo
  {journal} {Phys. Rev. B}\ }\textbf {\bibinfo {volume} {91}},\ \bibinfo
  {pages} {214309} (\bibinfo {year} {2015})},\ \Eprint
  {https://arxiv.org/abs/1511.07065} {1511.07065} \BibitemShut {NoStop}%
\bibitem [{\citenamefont {Plompen}\ \emph {et~al.}(2020)\citenamefont {Plompen}
  \emph {et~al.}}]{plompen:2020}%
  \BibitemOpen
  \bibfield  {author} {\bibinfo {author} {\bibfnamefont {A.~J.~M.}\
  \bibnamefont {Plompen}} \emph {et~al.},\ }\bibfield  {title} {\bibinfo
  {title} {The joint evaluated fission and fusion nuclear data library,
  {JEFF}-3.3},\ }\href {https://doi.org/10.1140/epja/s10050-020-00141-9}
  {\bibfield  {journal} {\bibinfo  {journal} {The European Physical Journal A}\
  }\textbf {\bibinfo {volume} {56}},\ \bibinfo {pages} {181} (\bibinfo {year}
  {2020})}\BibitemShut {NoStop}%
\bibitem [{\citenamefont {Otuka}\ \emph {et~al.}(2014)\citenamefont {Otuka}
  \emph {et~al.}}]{iaea:2014}%
  \BibitemOpen
  \bibfield  {author} {\bibinfo {author} {\bibfnamefont {N.}~\bibnamefont
  {Otuka}} \emph {et~al.},\ }\bibfield  {title} {\bibinfo {title} {Towards a
  more complete and accurate experimental nuclear reaction data library
  ({EXFOR}): {I}nternational collaboration between nuclear reaction data
  centres ({NRDC})},\ }\href {https://doi.org/10.1016/j.nds.2014.07.065}
  {\bibfield  {journal} {\bibinfo  {journal} {Nuclear Data Sheets}\ }\textbf
  {\bibinfo {volume} {120}},\ \bibinfo {pages} {272} (\bibinfo {year}
  {2014})}\BibitemShut {NoStop}%
\bibitem [{\citenamefont {Hamermesh}\ and\ \citenamefont
  {Schwinger}(1946)}]{hamermesh:1946}%
  \BibitemOpen
  \bibfield  {author} {\bibinfo {author} {\bibfnamefont {M.}~\bibnamefont
  {Hamermesh}}\ and\ \bibinfo {author} {\bibfnamefont {J.}~\bibnamefont
  {Schwinger}},\ }\bibfield  {title} {\bibinfo {title} {The {S}cattering of
  {S}low {N}eutrons by {O}rtho- and {P}aradeuterium},\ }\href@noop {}
  {\bibfield  {journal} {\bibinfo  {journal} {Phys. Rev.}\ }\textbf {\bibinfo
  {volume} {69}},\ \bibinfo {pages} {145} (\bibinfo {year} {1946})}\BibitemShut
  {NoStop}%
\bibitem [{\citenamefont {Young}\ and\ \citenamefont
  {Koppel}(1964)}]{young-koppel:1964}%
  \BibitemOpen
  \bibfield  {author} {\bibinfo {author} {\bibfnamefont {J.~A.}\ \bibnamefont
  {Young}}\ and\ \bibinfo {author} {\bibfnamefont {J.~U.}\ \bibnamefont
  {Koppel}},\ }\bibfield  {title} {\bibinfo {title} {{S}low {N}eutron
  {S}cattering by {M}olecular {H}ydrogen and {D}euterium},\ }\href
  {https://doi.org/10.1103/PhysRev.135.A603} {\bibfield  {journal} {\bibinfo
  {journal} {Phys. Rev.}\ }\textbf {\bibinfo {volume} {135}},\ \bibinfo {pages}
  {A603} (\bibinfo {year} {1964})}\BibitemShut {NoStop}%
\bibitem [{\citenamefont {Hill}\ and\ \citenamefont
  {Lounasmaa}(1959)}]{hill:1959}%
  \BibitemOpen
  \bibfield  {author} {\bibinfo {author} {\bibfnamefont {R.~W.}\ \bibnamefont
  {Hill}}\ and\ \bibinfo {author} {\bibfnamefont {O.~V.}\ \bibnamefont
  {Lounasmaa}},\ }\bibfield  {title} {\bibinfo {title} {The lattice specific
  heats of solid hydrogen and deuterium},\ }\href
  {https://doi.org/10.1080/14786435908238235} {\bibfield  {journal} {\bibinfo
  {journal} {Philos. Mag.}\ }\textbf {\bibinfo {volume} {4}},\ \bibinfo {pages}
  {785} (\bibinfo {year} {1959})}\BibitemShut {NoStop}%
\bibitem [{\citenamefont {Nielsen}(1973)}]{nielsen:1973}%
  \BibitemOpen
  \bibfield  {author} {\bibinfo {author} {\bibfnamefont {M.}~\bibnamefont
  {Nielsen}},\ }\bibfield  {title} {\bibinfo {title} {Phonons in solid hydrogen
  and deuterium studied by inelastic coherent neutron scattering},\ }\href
  {https://doi.org/10.1103/PhysRevB.7.1626} {\bibfield  {journal} {\bibinfo
  {journal} {Phys. Rev. B}\ }\textbf {\bibinfo {volume} {7}},\ \bibinfo {pages}
  {1626} (\bibinfo {year} {1973})}\BibitemShut {NoStop}%
\bibitem [{\citenamefont {Guarini}\ \emph {et~al.}(2016)\citenamefont
  {Guarini}, \citenamefont {Neumann}, \citenamefont {Bafile}, \citenamefont
  {Celli}, \citenamefont {Colognesi}, \citenamefont {Bellissima}, \citenamefont
  {Farhi},\ and\ \citenamefont {Calzavara}}]{guarini:2016}%
  \BibitemOpen
  \bibfield  {author} {\bibinfo {author} {\bibfnamefont {E.}~\bibnamefont
  {Guarini}}, \bibinfo {author} {\bibfnamefont {M.}~\bibnamefont {Neumann}},
  \bibinfo {author} {\bibfnamefont {U.}~\bibnamefont {Bafile}}, \bibinfo
  {author} {\bibfnamefont {M.}~\bibnamefont {Celli}}, \bibinfo {author}
  {\bibfnamefont {D.}~\bibnamefont {Colognesi}}, \bibinfo {author}
  {\bibfnamefont {S.}~\bibnamefont {Bellissima}}, \bibinfo {author}
  {\bibfnamefont {E.}~\bibnamefont {Farhi}},\ and\ \bibinfo {author}
  {\bibfnamefont {Y.}~\bibnamefont {Calzavara}},\ }\bibfield  {title} {\bibinfo
  {title} {{V}elocity autocorrelation by quantum simulations for direct
  parameter-free computations of the neutron cross sections. {II}. liquid
  deuterium},\ }\href {https://doi.org/10.1103/PhysRevB.93.224302} {\bibfield
  {journal} {\bibinfo  {journal} {Phys. Rev. B}\ }\textbf {\bibinfo {volume}
  {93}},\ \bibinfo {pages} {224302} (\bibinfo {year} {2016})}\BibitemShut
  {NoStop}%
\bibitem [{\citenamefont {Guarini}(2015)}]{guarini:2015-privcomm}%
  \BibitemOpen
  \bibfield  {author} {\bibinfo {author} {\bibfnamefont {E.}~\bibnamefont
  {Guarini}},\ }\href@noop {} {\bibfield  {journal} {\bibinfo  {journal}
  {{U}niversit\`a degli {S}tudi di {F}irenze, {I}taly}\ } (\bibinfo {year}
  {2015})},\ \bibinfo {note} {priv. comm.}\BibitemShut {Stop}%
\bibitem [{\citenamefont {Guarini}\ \emph {et~al.}(2015)\citenamefont
  {Guarini}, \citenamefont {Neumann}, \citenamefont {Bafile}, \citenamefont
  {Celli}, \citenamefont {Colognesi}, \citenamefont {Farhi},\ and\
  \citenamefont {Calzavara}}]{guarini:2015}%
  \BibitemOpen
  \bibfield  {author} {\bibinfo {author} {\bibfnamefont {E.}~\bibnamefont
  {Guarini}}, \bibinfo {author} {\bibfnamefont {M.}~\bibnamefont {Neumann}},
  \bibinfo {author} {\bibfnamefont {U.}~\bibnamefont {Bafile}}, \bibinfo
  {author} {\bibfnamefont {M.}~\bibnamefont {Celli}}, \bibinfo {author}
  {\bibfnamefont {D.}~\bibnamefont {Colognesi}}, \bibinfo {author}
  {\bibfnamefont {E.}~\bibnamefont {Farhi}},\ and\ \bibinfo {author}
  {\bibfnamefont {Y.}~\bibnamefont {Calzavara}},\ }\bibfield  {title} {\bibinfo
  {title} {Velocity autocorrelation in liquid parahydrogen by quantum
  simulations for direct parameter-free computations of neutron cross
  sections},\ }\href {https://doi.org/10.1103/PhysRevB.92.104303} {\bibfield
  {journal} {\bibinfo  {journal} {Phys. Rev. B}\ }\textbf {\bibinfo {volume}
  {92}},\ \bibinfo {pages} {104303} (\bibinfo {year} {2015})}\BibitemShut
  {NoStop}%
\bibitem [{\citenamefont {Bouguer}(1729)}]{bouguer:1729}%
  \BibitemOpen
  \bibfield  {author} {\bibinfo {author} {\bibfnamefont {P.}~\bibnamefont
  {Bouguer}},\ }\href@noop {} {\emph {\bibinfo {title} {Essai d'optique sur la
  gradation de la lumi\`ere}}}\ (\bibinfo  {publisher} {Claude Jombert,
  Paris},\ \bibinfo {year} {1729})\BibitemShut {NoStop}%
\bibitem [{\citenamefont {Beer}(1852)}]{beer:1852}%
  \BibitemOpen
  \bibfield  {author} {\bibinfo {author} {\bibfnamefont {A.}~\bibnamefont
  {Beer}},\ }\bibfield  {title} {\bibinfo {title} {{B}estimmung der
  {A}bsorption des rothen {L}ichts in farbigen {F}l\"{u}ssigkeiten},\
  }\href@noop {} {\bibfield  {journal} {\bibinfo  {journal} {Annalen der Physik
  und Chemie}\ }\textbf {\bibinfo {volume} {86}},\ \bibinfo {pages} {78}
  (\bibinfo {year} {1852})}\BibitemShut {NoStop}%
\bibitem [{\citenamefont {Steyerl}\ \emph {et~al.}(1986)\citenamefont
  {Steyerl}, \citenamefont {Nagel}, \citenamefont {Schreiber}, \citenamefont
  {Steinhauser}, \citenamefont {G\"{a}hler}, \citenamefont {Gl\"{a}ser},
  \citenamefont {Ageron}, \citenamefont {Astruc}, \citenamefont {Drexel},
  \citenamefont {Gervais},\ and\ \citenamefont {Mampe}}]{steyerl:1986}%
  \BibitemOpen
  \bibfield  {author} {\bibinfo {author} {\bibfnamefont {A.}~\bibnamefont
  {Steyerl}}, \bibinfo {author} {\bibfnamefont {H.}~\bibnamefont {Nagel}},
  \bibinfo {author} {\bibfnamefont {F.-X.}\ \bibnamefont {Schreiber}}, \bibinfo
  {author} {\bibfnamefont {K.-A.}\ \bibnamefont {Steinhauser}}, \bibinfo
  {author} {\bibfnamefont {R.}~\bibnamefont {G\"{a}hler}}, \bibinfo {author}
  {\bibfnamefont {W.}~\bibnamefont {Gl\"{a}ser}}, \bibinfo {author}
  {\bibfnamefont {P.}~\bibnamefont {Ageron}}, \bibinfo {author} {\bibfnamefont
  {J.~M.}\ \bibnamefont {Astruc}}, \bibinfo {author} {\bibfnamefont
  {W.}~\bibnamefont {Drexel}}, \bibinfo {author} {\bibfnamefont
  {G.}~\bibnamefont {Gervais}},\ and\ \bibinfo {author} {\bibfnamefont
  {W.}~\bibnamefont {Mampe}},\ }\bibfield  {title} {\bibinfo {title} {A new
  source of cold and ultracold neutrons},\ }\href
  {https://doi.org/10.1016/0375-9601(86)90587-6} {\bibfield  {journal}
  {\bibinfo  {journal} {Physics Letters A}\ }\textbf {\bibinfo {volume}
  {116}},\ \bibinfo {pages} {347} (\bibinfo {year} {1986})}\BibitemShut
  {NoStop}%
\bibitem [{\citenamefont {D\"{o}ge}\ \emph {et~al.}(2020)\citenamefont
  {D\"{o}ge}, \citenamefont {Hingerl},\ and\ \citenamefont
  {Morkel}}]{doege:2020-turbine}%
  \BibitemOpen
  \bibfield  {author} {\bibinfo {author} {\bibfnamefont {S.}~\bibnamefont
  {D\"{o}ge}}, \bibinfo {author} {\bibfnamefont {J.}~\bibnamefont {Hingerl}},\
  and\ \bibinfo {author} {\bibfnamefont {C.}~\bibnamefont {Morkel}},\
  }\bibfield  {title} {\bibinfo {title} {Measured velocity spectra and neutron
  densities of the {PF2} ultracold-neutron beam ports at the {I}nstitut
  {L}aue–{L}angevin},\ }\href {https://doi.org/10.1016/j.nima.2019.163112}
  {\bibfield  {journal} {\bibinfo  {journal} {Nuclear Instruments and Methods
  in Physics Research Section A: Accelerators, Spectrometers, Detectors and
  Associated Equipment}\ }\textbf {\bibinfo {volume} {953}},\ \bibinfo {pages}
  {163112} (\bibinfo {year} {2020})}\BibitemShut {NoStop}%
\bibitem [{\citenamefont {D\"oge}\ \emph {et~al.}(2020)\citenamefont {D\"oge},
  \citenamefont {Hingerl}, \citenamefont {Lychagin},\ and\ \citenamefont
  {Morkel}}]{doege:2020-foils}%
  \BibitemOpen
  \bibfield  {author} {\bibinfo {author} {\bibfnamefont {S.}~\bibnamefont
  {D\"oge}}, \bibinfo {author} {\bibfnamefont {J.}~\bibnamefont {Hingerl}},
  \bibinfo {author} {\bibfnamefont {E.~V.}\ \bibnamefont {Lychagin}},\ and\
  \bibinfo {author} {\bibfnamefont {C.}~\bibnamefont {Morkel}},\ }\bibfield
  {title} {\bibinfo {title} {Scattering of ultracold neutrons from rough
  surfaces of metal foils},\ }\href
  {https://doi.org/10.1103/PhysRevC.102.064607} {\bibfield  {journal} {\bibinfo
   {journal} {Phys. Rev. C}\ }\textbf {\bibinfo {volume} {102}},\ \bibinfo
  {pages} {064607} (\bibinfo {year} {2020})}\BibitemShut {NoStop}%
\bibitem [{\citenamefont {D\"{o}ge}\ and\ \citenamefont
  {Hingerl}(2018)}]{doege:2018}%
  \BibitemOpen
  \bibfield  {author} {\bibinfo {author} {\bibfnamefont {S.}~\bibnamefont
  {D\"{o}ge}}\ and\ \bibinfo {author} {\bibfnamefont {J.}~\bibnamefont
  {Hingerl}},\ }\bibfield  {title} {\bibinfo {title} {A hydrogen leak-tight,
  transparent cryogenic sample container for ultracold-neutron transmission
  measurements},\ }\href {https://doi.org/10.1063/1.4996296} {\bibfield
  {journal} {\bibinfo  {journal} {Review of Scientific Instruments}\ }\textbf
  {\bibinfo {volume} {89}},\ \bibinfo {pages} {033903} (\bibinfo {year}
  {2018})},\ \Eprint {https://arxiv.org/abs/1803.10159} {1803.10159}
  \BibitemShut {NoStop}%
\bibitem [{\citenamefont {D\"{o}ge}(2019)}]{doege:2019-phd}%
  \BibitemOpen
  \bibfield  {author} {\bibinfo {author} {\bibfnamefont {S.}~\bibnamefont
  {D\"{o}ge}},\ }\emph {\bibinfo {title} {{S}cattering of {U}ltracold
  {N}eutrons in {C}ondensed {D}euterium and on {M}aterial {S}urfaces}},\ \href
  {https://doi.org/10.14459/2019md1464401} {\bibinfo {type} {{Ph.D.} thesis}},\
  \bibinfo  {school} {Technische {U}niversit\"{a}t M\"{u}nchen, Munich,
  Germany} (\bibinfo {year} {2019}),\ \bibinfo {note}
  {http://doi.org/10.14459/2019md1464401}\BibitemShut {NoStop}%
\bibitem [{\citenamefont {Doege}\ \emph {et~al.}(2017)\citenamefont {Doege},
  \citenamefont {Morkel}, \citenamefont {Hingerl}, \citenamefont {Lauss},\ and\
  \citenamefont {Hild}}]{doege:2017-3-14-374}%
  \BibitemOpen
  \bibfield  {author} {\bibinfo {author} {\bibfnamefont {S.}~\bibnamefont
  {Doege}}, \bibinfo {author} {\bibfnamefont {C.}~\bibnamefont {Morkel}},
  \bibinfo {author} {\bibfnamefont {J.}~\bibnamefont {Hingerl}}, \bibinfo
  {author} {\bibfnamefont {B.}~\bibnamefont {Lauss}},\ and\ \bibinfo {author}
  {\bibfnamefont {N.}~\bibnamefont {Hild}},\ }\bibfield  {title} {\bibinfo
  {title} {Measurement of the mean free path of ultracold neutrons ({UCN}s) in
  liquid and solid hydrogen and deuterium at various temperatures},\ }\bibfield
   {journal} {\bibinfo  {journal} {Institut Laue-Langevin}\ }\href
  {https://doi.org/10.5291/ill-data.3-14-374} {10.5291/ill-data.3-14-374}
  (\bibinfo {year} {2017}),\ \bibinfo {note}
  {https://doi.org/10.5291/ill-data.3-14-374}\BibitemShut {NoStop}%
\bibitem [{\citenamefont {Klein}\ and\ \citenamefont
  {Schmidt}(2011)}]{klein:2011}%
  \BibitemOpen
  \bibfield  {author} {\bibinfo {author} {\bibfnamefont {M.}~\bibnamefont
  {Klein}}\ and\ \bibinfo {author} {\bibfnamefont {C.~J.}\ \bibnamefont
  {Schmidt}},\ }\bibfield  {title} {\bibinfo {title} {{CASCADE}, neutron
  detectors for highest count rates in combination with {ASIC}/{FPGA} based
  readout electronics},\ }\href {https://doi.org/10.1016/j.nima.2010.06.278}
  {\bibfield  {journal} {\bibinfo  {journal} {Nuclear Instruments and Methods
  in Physics Research Section A: Accelerators, Spectrometers, Detectors and
  Associated Equipment}\ }\textbf {\bibinfo {volume} {628}},\ \bibinfo {pages}
  {9} (\bibinfo {year} {2011})},\ \bibinfo {note} {{VCI} 2010 Proceedings of
  the 12th International Vienna Conference on Instrumentation}\BibitemShut
  {NoStop}%
\bibitem [{\citenamefont {Sears}(1992)}]{sears:1992}%
  \BibitemOpen
  \bibfield  {author} {\bibinfo {author} {\bibfnamefont {V.~F.}\ \bibnamefont
  {Sears}},\ }\bibfield  {title} {\bibinfo {title} {Neutron scattering lengths
  and cross sections},\ }\href {http://www.ncnr.nist.gov/resources/n-lengths/}
  {\bibfield  {journal} {\bibinfo  {journal} {Neutron News}\ }\textbf {\bibinfo
  {volume} {3}},\ \bibinfo {pages} {26} (\bibinfo {year} {1992})}\BibitemShut
  {NoStop}%
\bibitem [{\citenamefont {Liu}\ \emph {et~al.}(2000)\citenamefont {Liu},
  \citenamefont {Young},\ and\ \citenamefont {Lamoreaux}}]{liu:2000}%
  \BibitemOpen
  \bibfield  {author} {\bibinfo {author} {\bibfnamefont {C.-Y.}\ \bibnamefont
  {Liu}}, \bibinfo {author} {\bibfnamefont {A.~R.}\ \bibnamefont {Young}},\
  and\ \bibinfo {author} {\bibfnamefont {S.~K.}\ \bibnamefont {Lamoreaux}},\
  }\bibfield  {title} {\bibinfo {title} {Ultracold neutron upscattering rates
  in a molecular deuterium crystal},\ }\href
  {https://doi.org/10.1103/PhysRevB.62.R3581} {\bibfield  {journal} {\bibinfo
  {journal} {Phys. Rev. B}\ }\textbf {\bibinfo {volume} {62}},\ \bibinfo
  {pages} {R3581} (\bibinfo {year} {2000})}\BibitemShut {NoStop}%
\bibitem [{\citenamefont {O'Reilly}\ and\ \citenamefont
  {Peterson}(1977)}]{oreilly:1977}%
  \BibitemOpen
  \bibfield  {author} {\bibinfo {author} {\bibfnamefont {D.~E.}\ \bibnamefont
  {O'Reilly}}\ and\ \bibinfo {author} {\bibfnamefont {E.~M.}\ \bibnamefont
  {Peterson}},\ }\bibfield  {title} {\bibinfo {title} {Self-diffusion of liquid
  hydrogen and deuterium},\ }\href {https://doi.org/10.1063/1.434001}
  {\bibfield  {journal} {\bibinfo  {journal} {The Journal of Chemical Physics}\
  }\textbf {\bibinfo {volume} {66}},\ \bibinfo {pages} {934} (\bibinfo {year}
  {1977})}\BibitemShut {NoStop}%
\bibitem [{\citenamefont {Souers}(1986)}]{souers:1986}%
  \BibitemOpen
  \bibfield  {author} {\bibinfo {author} {\bibfnamefont {P.~C.}\ \bibnamefont
  {Souers}},\ }\href@noop {} {\emph {\bibinfo {title} {Hydrogen Properties for
  Fusion Energy}}}\ (\bibinfo  {publisher} {University of California Press,
  USA},\ \bibinfo {year} {1986})\BibitemShut {NoStop}%
\bibitem [{\citenamefont {D\"{o}ge}\ \emph {et~al.}(2016)\citenamefont
  {D\"{o}ge}, \citenamefont {Herold}, \citenamefont {Gutsmiedl}, \citenamefont
  {Morkel},\ and\ \citenamefont {Geltenbort}}]{doege:2016}%
  \BibitemOpen
  \bibfield  {author} {\bibinfo {author} {\bibfnamefont {S.}~\bibnamefont
  {D\"{o}ge}}, \bibinfo {author} {\bibfnamefont {C.}~\bibnamefont {Herold}},
  \bibinfo {author} {\bibfnamefont {E.}~\bibnamefont {Gutsmiedl}}, \bibinfo
  {author} {\bibfnamefont {C.}~\bibnamefont {Morkel}},\ and\ \bibinfo {author}
  {\bibfnamefont {P.}~\bibnamefont {Geltenbort}},\ }\bibfield  {title}
  {\bibinfo {title} {New experimental results for the scattering cross sections
  of liquid and solid deuterium for ultracold neutrons and an approach to their
  calculation},\ }\href {http://isinn.jinr.ru/proceedings/isinn-23.html}
  {\bibfield  {journal} {\bibinfo  {journal} {ISINN Conference Proceedings}\
  }\textbf {\bibinfo {volume} {23}},\ \bibinfo {pages} {119} (\bibinfo {year}
  {2016})}\BibitemShut {NoStop}%
\bibitem [{\citenamefont {Turchin}(1965)}]{turchin:1965}%
  \BibitemOpen
  \bibfield  {author} {\bibinfo {author} {\bibfnamefont {V.~F.}\ \bibnamefont
  {Turchin}},\ }\href@noop {} {\emph {\bibinfo {title} {Slow Neutrons}}},\
  {I}srael program for scientific translations\ (\bibinfo  {publisher} {Sivan
  Press, Jerusalem},\ \bibinfo {year} {1965})\ \bibinfo {note} {{R}ussian
  original: Medlennye nejtrony (Gosatomizdat, Moscow, 1963)}\BibitemShut
  {NoStop}%
\bibitem [{\citenamefont {D\"oge}\ \emph {et~al.}(2021)\citenamefont {D\"oge},
  \citenamefont {Liu}, \citenamefont {Young},\ and\ \citenamefont
  {Morkel}}]{doege:2021-inc-approx}%
  \BibitemOpen
  \bibfield  {author} {\bibinfo {author} {\bibfnamefont {S.}~\bibnamefont
  {D\"oge}}, \bibinfo {author} {\bibfnamefont {C.-Y.}\ \bibnamefont {Liu}},
  \bibinfo {author} {\bibfnamefont {A.}~\bibnamefont {Young}},\ and\ \bibinfo
  {author} {\bibfnamefont {C.}~\bibnamefont {Morkel}},\ }\bibfield  {title}
  {\bibinfo {title} {Incoherent approximation for neutron up-scattering cross
  sections and its corrections for slow neutrons and low crystal
  temperatures},\ }\href {https://doi.org/10.1103/PhysRevC.103.054606}
  {\bibfield  {journal} {\bibinfo  {journal} {Phys. Rev. C}\ }\textbf {\bibinfo
  {volume} {103}},\ \bibinfo {pages} {054606} (\bibinfo {year}
  {2021})}\BibitemShut {NoStop}%
\bibitem [{\citenamefont {Bernnat}\ \emph {et~al.}(2002)\citenamefont
  {Bernnat}, \citenamefont {Keinert},\ and\ \citenamefont
  {Mattes}}]{bernnat:2002}%
  \BibitemOpen
  \bibfield  {author} {\bibinfo {author} {\bibfnamefont {W.}~\bibnamefont
  {Bernnat}}, \bibinfo {author} {\bibfnamefont {J.}~\bibnamefont {Keinert}},\
  and\ \bibinfo {author} {\bibfnamefont {M.}~\bibnamefont {Mattes}},\
  }\bibfield  {title} {\bibinfo {title} {Scattering laws and cross sections for
  moderators and structure materials for calculation of production and
  transport of cold and ultracold neutrons},\ }\href
  {https://doi.org/10.1080/00223131.2002.10875056} {\bibfield  {journal}
  {\bibinfo  {journal} {Journal of Nuclear Science and Technology}\ }\textbf
  {\bibinfo {volume} {39}},\ \bibinfo {pages} {124} (\bibinfo {year}
  {2002})}\BibitemShut {NoStop}%
\end{thebibliography}

%

\end{document}